\documentclass[reprint,aps,prmaterials,amsmath,amssymb,superscriptaddress,floatfix]{revtex4-2}

\usepackage{lmodern}
\usepackage{graphicx}
\usepackage{xcolor}
\usepackage{soul}
\usepackage{booktabs}

\usepackage{hyperref}
\hypersetup{colorlinks=true,allcolors=blue,pdfstartview={FitH}}
\usepackage[all]{hypcap}

\newcommand{\tuwien}{Institute of Applied Physics, TU Wien, Wiedner Hauptstra{\ss}e 8-10/E134, A-1040 Wien, Austria}
\newcommand{\tumunich}{Department of Chemistry, TUM School of Natural Sciences, Technical University of Munich, Lichtenbergstra{\ss}e 4, D-85748 Garching b.\ M\"unchen, Germany}
\newcommand{\fauerlangen}{Institute of Condensed Matter Physics, Universit\"at Erlangen-N\"urnberg, Staudtstra{\ss}e 7, D-91058 Erlangen, Germany}

\DeclareRobustCommand{\hematite}{\texorpdfstring{$\alpha\text{-Fe}_2\text{O}_3(1\overline{1}02)$}{a-Fe2O3(1-102)}}
\DeclareRobustCommand{\hematiteOneby}{\hematite{}-\woods{1}{1}{}}
\DeclareRobustCommand{\iv}{\texorpdfstring{$I(V)$}{I(V)}}
\DeclareRobustCommand{\leediv}{LEED~\iv{}}
\DeclareRobustCommand{\leedivdash}{LEED-\iv{}}
\DeclareRobustCommand{\deltas}{$\delta{}A_{i,\mathbf{g}}$}
\DeclareRobustCommand{\program}[1]{\textsc{#1}}
\DeclareRobustCommand{\tenserleed}{\program{TensErLEED}}
\DeclareRobustCommand{\python}{\program{Python}}
\DeclareRobustCommand{\calc}{\program{viperleed.calc}}
\DeclareRobustCommand{\ViPErLEED}{ViPErLEED}
\DeclareRobustCommand{\fortran}{\program{Fortran}}
\DeclareRobustCommand{\param}[1]{\texttt{#1}}
\DeclareRobustCommand{\lmax}{$\ell_\text{max}$}
\DeclareRobustCommand{\SI}{Supplemental Material \cite{supporting}}
\DeclareRobustCommand{\viperSpot}{Part~II \cite{viperleedSpot}}
\DeclareRobustCommand{\viperMeas}{Part~III \cite{viperleedMeasurement}}
\DeclareRobustCommand{\viperDoc}{\ViPErLEED{} documentation \cite{viperleedOrg}}

\DeclareRobustCommand{\woods}[3]{%
\texorpdfstring{\mbox{$(#1\hspace{0.1em}\times\hspace{0.1em}#2)$}}{(#1x#2)}%
\if\relax\detokenize{#3}\relax%
\relax%
\else%
\texorpdfstring{$R$#3$^\circ$}{R#3}%
\fi
}

\DeclareRobustCommand{\degC}{$^\circ$C}
\DeclareRobustCommand{\quotes}[1]{``#1''}
\newlength{\pagefigure}
\newlength{\columnfigure}
\DeclareRobustCommand{\floatref}[3][]{%
\hyperref[#2]{#3}~\ref{#2}%
\if\relax\detokenize{#1}\relax%
\relax%
\else%
\hyperref[#2]{(#1)}%
\fi}

\DeclareRobustCommand{\Fig}[2][]{\floatref[#1]{#2}{Fig.}}
\DeclareRobustCommand{\Figure}[2][]{\floatref[#1]{#2}{Figure}}
\DeclareRobustCommand{\Table}[1]{\floatref{#1}{Table}}

\DeclareRobustCommand{\Section}[1]{\hyperref[#1]{Section}~\ref{#1}}
\DeclareRobustCommand{\RefInline}[1]{Ref.~\onlinecite{#1}}
\DeclareRobustCommand{\RefsInline}[1]{Refs.~\onlinecite{#1}}

\DeclareRobustCommand{\todo}[1]{\sethlcolor{yellow}\hl{#1}}%
\DeclareRobustCommand{\todoHidden}[1]{}

\AtBeginDocument{%
    \newwrite\bibnotes
    \def\bibnotesext{Notes.bib}
    \immediate\openout\bibnotes=\jobname\bibnotesext
    \immediate\write\bibnotes{@CONTROL{REVTEX42Control}}
    \immediate\write\bibnotes{@CONTROL{%
    apsrev42Control,author="08",editor="1",pages="1",title="0",year="1"}}
     \if@filesw
     \immediate\write\@auxout{\string\citation{apsrev42Control}}%
    \fi
}%

\begin{document}

\title{\boldmath{}\ViPErLEED{} package I: %
       Calculation of \iv{} curves and structural optimization}
\date{\today}

\author{Florian Kraushofer}
\affiliation{\tuwien}
\affiliation{\tumunich}
\author{Alexander M. Imre}
\author{Giada Franceschi}
\affiliation{\tuwien}
\author{Tilman Ki{\ss}linger}
\affiliation{\fauerlangen}
\author{Erik Rheinfrank}
\author{Michael Schmid}
\author{Ulrike Diebold}
\affiliation{\tuwien}
\author{Lutz Hammer}
\affiliation{\fauerlangen}
\author{Michele Riva}
\email[Corresponding author: ]{riva@iap.tuwien.ac.at}
\affiliation{\tuwien}

\begin{abstract}
Low-energy electron diffraction (LEED) is a widely used technique in surface-science laboratories. Yet, it is rarely used to its full potential. The quantitative information about the surface structure, contained in the modulation of the intensities of the diffracted beams as a function of incident electron energy, \leediv{}, is underutilized. To acquire these data, only minor adjustments would be required in most experimental setups, but existing analysis software is cumbersome to use and often computationally inefficient. The \ViPErLEED{} (Vienna package for Erlangen LEED) project lowers these barriers, introducing a combined solution for user-friendly data acquisition, extraction, and computational analysis. These parts are discussed in three separate publications. Here, the focus is on the computational part of \ViPErLEED{}, which performs highly automated \leedivdash{} calculations and structural optimization. Minimal user input is required, and the functionality is significantly enhanced compared to existing solutions. Computation is performed by embedding the existing Erlangen tensor-LEED package (\tenserleed{}). \ViPErLEED{} manages additional parallelization, monitors convergence, and processes all input and output. This makes \leediv{} more accessible to new users while minimizing the potential for errors and the manual labor. Added functionalities include intelligent structure-dependent defaults for most calculation parameters, automatic detection of bulk and surface symmetries and their relationship, automated search procedures that preserve the symmetry and speed up convergence, adjustments to the \tenserleed{} code to handle larger systems than before, as well as parallelization and optimization. Modern file formats are used as input and output, and there is a direct interface to the Atomic Simulation Environment (ASE) package. The software is implemented primarily in \python{} (version $\geq3.7$) and provided as an open-source package (GNU GPLv3 or any later version). A structure determination of the \hematiteOneby{} surface is presented as an example for the application of the software.
\end{abstract}

\maketitle

\section{Introduction}

\subsection{\boldmath{}History and motivation: The untapped potential of \leediv{}}
Low-energy electron diffraction (LEED) is a versatile, quick and easy-to-use characterization technique routinely used in many surface-science laboratories. Acquiring LEED images typically takes minutes, limited more by the time it takes to move a sample into position than the actual measurement. The resulting LEED pattern offers a reciprocal-space fingerprint of the surface unit cell, where each \quotes{spot} corresponds to a beam of diffracted electrons that fulfills the constructive-interference Bragg condition of the 2D periodicity of the surface. These data can indicate whether a given sample preparation yields the desired surface structure. However, this snapshot of the reciprocal lattice is only a fraction of the full information available from LEED. Spot-profile analysis (SPA)LEED \cite{spaleed1986,spaleed2024}, as well as analysis of the background \cite{dleed1996}, can offer information about defects and surface order. \leediv{} can yield accurate positions for the atoms in the unit cell. In \leediv{}, the intensities $I$ of the scattered electron beams are measured as a function of electron energy, tuned via the accelerating voltage $V$. Since these \quotes{\iv{} curves} can also be calculated theoretically, experimental \leedivdash{} data can be directly compared to a structural model of a surface. Comparison between calculations and experiment is performed through the so-called $R$ factor, which evaluates agreement across the entire beam set and breaks it down into a single value \cite{pendry1980rfactor,moore1982rfactors}. Thus, with a relatively quick and simple laboratory-based experiment, one can confirm or reject hypothetical surface models, perform local optimizations of the atomic coordinates in such a model, or obtain a robust fingerprint to confirm the reproducibility of sample preparation \cite{moritzVanhove2022,heinz2013book}.

\begin{figure*}[t]
\includegraphics[width=0.8\pagefigure]{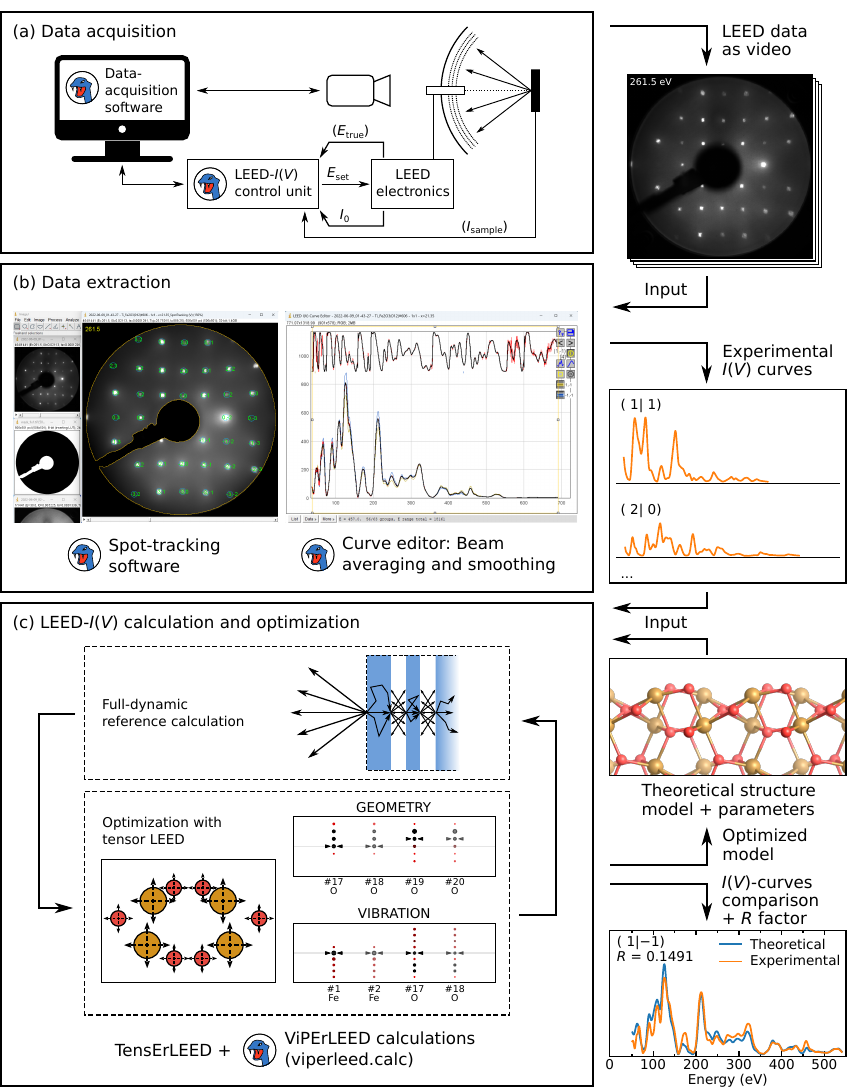}
\caption{\label{fig:overview}Overview of the \ViPErLEED{} package. Data acquisition (a) and data extraction (b) are discussed in \viperMeas{} and \viperSpot{}, respectively. This work describes the \leedivdash{} calculations, indicated in panel (c). The relationship between \tenserleed{} and \ViPErLEED{} is visualized in more detail in \Fig{fig:workflow}. Example data in this figure are from measurements on \hematiteOneby{}.}
\end{figure*}

Crucially, the hardware for these experiments is already present in many surface-science laboratories, and only minor adjustments are required to upgrade most standard LEED setups to \leedivdash{} capability. The bigger hurdles are the analysis of the data and the calculation of the theoretical \iv{} curves based on a given model. \ViPErLEED{}, presented here, is a comprehensive package that combines a simple hardware add-on with software for acquiring and extracting the \iv{} data, and for calculating and comparing to theoretical \iv{} curves. \Figure{fig:overview} shows the workflow with input and output.

The whole package consists of three parts, which are presented in three separate publications: The data acquisition (hardware), discussed in \viperMeas{}; the data extraction (spot tracking) and processing, discussed in \viperSpot{}; and the \leedivdash{} calculation and optimization, presented here. Specifically, this work discusses the software to calculate \iv{} curves for a given reference structure and to optimize this structure based on the match with experimental data. While these three parts are optimized to be used together, they all employ standard file formats (such as comma-separated values --- CSV) for input and output and can be used independently.

\subsection{\boldmath{}Basics of \leediv{}\label{sec:leed_theory}}
An extensive treatment of the theory of quantitative LEED, as well as the standard computational treatment, can be found in \RefsInline{vanhovetong1979,pendry1980rfactor,rous1986prl,rous1989,rous1992,tenserleed,heinz2013book,fauster2020,pendry1974,vanhove1986,moritzVanhove2022}. These are only outlined briefly here to introduce important terms and concepts used in this work.

LEED is highly surface sensitive due to the low inelastic mean free path of electrons in solids ($\approx$5--15~\AA{} in the relevant energy range of up to 1000~eV \cite{seah1979}). Because of the strong electron--solid interaction, electrons typically experience multiple scattering events before leaving the surface \cite{foot_info_in_iv}. Thus, calculations of \leedivdash{} spectra need to include paths with multiple scattering events. Crucially, small displacements (on the order of a few picometres) strongly affect the intensities of scattered beams. %
The high sensitivity to atomic positions gives \leediv{} the potential for accurate determination of atomic coordinates. This, however, can only be achieved through a careful optimization process while comparing calculated and experimental \iv{} curves. 

\begin{figure}[b]
\includegraphics[width=\columnfigure]{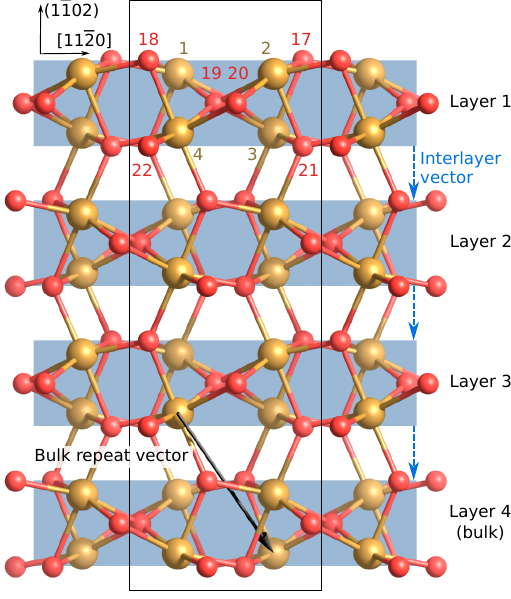}
\caption{\label{fig:fe2o3:layers}Example of layer definitions as used in \ViPErLEED{} and \tenserleed{}, shown for the \hematiteOneby{} bulk-truncated surface (side view along $[\overline{1}101]$) \cite{henderson2002,lad1988,tanwar2007,franceschi2020,kraushofer2018}. The layers are indicated by blue rectangles. Each layer contains four Fe atoms (large yellow spheres) and six O atoms (small red spheres) per unit cell. Interlayer vectors as used in \tenserleed{} are drawn as dashed blue arrows. Note that the interlayer vectors do not connect layer origins, but rather point from the $z$ coordinate of the bottommost atom in one layer to the $z$ coordinate of the topmost atom in the subsequent underlying layer. Layer~4 is defined as bulk. The black arrow indicates the bulk repeat vector, which can be automatically detected by \ViPErLEED{} (see \Section{sec:find_bulk}).}
\end{figure}

Scattering of low-energy ($\approx30$--2000~eV) electrons with one atom in a solid is well approximated as occurring in a spherically symmetric potential (muffin-tin approximation) and conserving the angular momentum $\ell$. Single scattering events of electrons with energy $E$ at a given site can then be described by partial-wave phase shifts $\delta_\ell(E)$. Electron--phonon scattering is included as an effective Debye--Waller factor, which depends on the vibration amplitudes of the atoms. Inelastic scattering is treated via an imaginary part of the inner potential, leading to an exponential decay with distance. The real part of the inner potential is defined by an approximate function, derived together with the scattering phase shifts with a boundary condition to the vacuum \cite{rundgren2003,rundgren2007}. With these few approximations, one can perform full-dynamic calculations of the multiple scattering of electrons, yielding theoretical \iv{} curves that can be compared to experiment \cite{vanhovetong1979}.

In a typical LEED calculation, the surface is split into \quotes{layers}, which should be separated along the surface normal by distances of at least $1$~\AA{}. Each layer can consist of one or multiple atoms. Multiple scattering within each layer is then evaluated in angular-momentum space, resulting in layer-diffraction matrices. Propagation and scattering between the layers can be treated in a plane-wave representation \cite{tenserleed,vanhovetong1979}. \Figure{fig:fe2o3:layers} shows a typical layer definition for an example system of medium complexity: the \hematiteOneby{} surface \cite{kraushofer2018}. This system will be referred to several times below, and it is one of the examples given in the \ViPErLEED{} documentation (see the \SI{}). This layering scheme also allows for straightforward treatment of scattering far below the surface: One or two layers are defined as a bulk repeat unit (e.g., Layer~4 in \Fig{fig:fe2o3:layers}) with a given bulk repeat vector. This bulk unit can then be stacked through repeated doubling until the norm of the reflection and transmission matrices of the bulk remains unchanged between successive steps \cite{tenserleed}.

\subsection{The tensor-LEED approximation\label{sec:tensor_leed}}

\begin{figure*}[tb]
\includegraphics[width=.7\pagefigure]{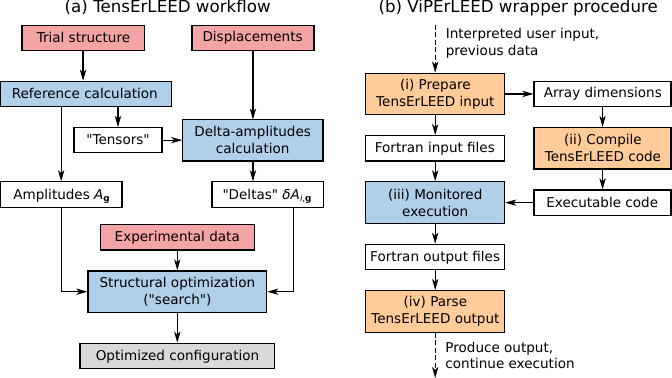}
\caption{\label{fig:workflow}(a) Simplified sketch of the \tenserleed{} workflow as described in \RefInline{tenserleed}. Red boxes represent user input (in \tenserleed{} formats), blue boxes represent the main \tenserleed{} modules. Additional modules used to calculate the $R$ factor, produce theoretical beams after the search, or perform error calculations are not drawn. (b) Sketch showing how \tenserleed{} modules are embedded into the \ViPErLEED{} code. Orange boxes represent logical steps, the blue box (iii) can stand for any \tenserleed{} module.}
\end{figure*}

The \quotes{full-dynamic} LEED calculations discussed so far are computationally demanding. The most expensive part is the inversion of matrices needed for intralayer scattering, which scales as $d^n$ ($n=2.38$--3, depending on the algorithm) for $d\times{}d$ matrices. Intralayer scattering matrices have $d=N(\ell_\text{max}+1)^2$ for $N$ atoms in the layer and a maximum angular momentum \lmax{} \cite{heinz2013book}.  As discussed in \Section{sec:leed_theory}, the high sensitivity to atomic coordinates on the picometre scale implies that basic comparisons of theoretical models to experimental data require local optimization of these coordinates, entailing numerous calculations. The number of optimization parameters also scales with system size, with a concomitant increase in the number of trial calculations \cite{kottcke1997}. 

The so-called tensor-LEED approximation makes these optimizations more efficient: A full-dynamic calculation is performed only for a reference structure; changes to the diffraction intensities through small modifications of the reference structure are treated by first-order perturbation theory \cite{rous1986prl,rous1989,rous1992}. This approach is ideal for structural optimization aimed at improving the agreement between theoretical and experimental \iv{} curves, as the full-dynamic calculation only needs to be performed once. The optimization can then be performed using much cheaper evaluations of a set of perturbations, which scale linearly with the number of parameters under variation \cite{rous1989}. Importantly, the initial structural guess must be close to the target structure: The tensor-LEED approximation breaks down for displacements of more than $\approx0.4$~\AA{} \cite{rous1989}, and significant errors are already introduced much sooner. This means that new reference calculations must be performed when changes to the original structure become too large. It is worth noting that the tensor-LEED approximation is not limited to purely geometrical perturbations of the reference structure. Changes of atomic vibration amplitudes \cite{loffler1994} and statistical chemical substitutions \cite{doll1993} (including vacancies) can be treated with the same method and even concurrently \cite[p.~135]{heinz2013book}.

Several implementations exist for calculating \leedivdash{} curves, most of which are derived from the pioneering works of Pendry \cite{pendry1974}, Tong \cite{tong1975}, and Van~Hove \cite{vanhovetong1979}. Perhaps the most widely known is the Barbieri--Van Hove symmetrized automated LEED (SATLEED) package and its derivatives (MSATLEED, SATCLEED, ATLMLEED) \cite{satleed,vanhove_web}. Alternatives include \tenserleed{} \cite{tenserleed}, described below, as well as DL\_LEED \cite{wander2001}, LEED90 \cite{leed90}, CAVLEED \cite{cavleed}, CLEED \cite{cleed} and LEEDFIT \cite{rundgren2021}, though the latter four do not implement the tensor-LEED approximation, while DL\_LEED has a partial tensor-LEED implementation \cite{foot_more_leed_codes}. The various packages encompass different capabilities, but none automates the structure-optimization process. Recently, the AQuaLEED package was published to simplify use of the SATLEED codes through a \python{} wrapper program \cite{mayer2012,easyleed} in a similar, though much less broad, approach as the \ViPErLEED{} package described here. A \python{} wrapper has also been recently developed for CLEED \cite{cleedpy}.

The theoretical foundation and code architecture of the \tenserleed{} package are described in detail in \RefInline{tenserleed}. Briefly, the package consists of three main modules, shown in \Fig[a]{fig:workflow}, which perform different segments of the \leedivdash{} calculation. The \quotes{reference-calculation} module performs full-dynamic \leedivdash{} calculations, producing the complex amplitudes $A_\mathbf{g}$ of the diffracted beams, $\mathbf{g}$, as implemented previously in the Van Hove--Tong code \cite{vanhovetong1979}. In addition, it also stores information necessary for the tensor-LEED approximation in separate output files, named \quotes{Tensors} in \RefInline{tenserleed}. The second \tenserleed{} segment is a \quotes{delta-amplitudes} calculation, which computes the changes \deltas{} for a small perturbation to the position, vibration amplitude, or site occupation (including mixing of chemical species as described by the average $t$-matrix approximation \cite{crampin1991}) of an atom $i$. These \deltas{} are calculated for all atoms under variation and for each step in a given atom's range of displacements. The third segment, the \quotes{search}, aims for the set of \deltas{} that, together with the reference amplitudes $A_\mathbf{g}$, minimizes the $R$ factor (i.e., the disagreement) between the calculated and experimental beams. Simply put, the full-dynamic reference calculation is a \quotes{precise} LEED calculation for a given reference structure, while the \deltas{} calculation and \quotes{search} compose the structural optimization using tensor LEED.

Altogether, the \tenserleed{} package \cite{tenserleed} is a powerful and versatile tool for \leedivdash{} calculations, but it has shortcomings. Employing the package with its custom settings demands significant experience and poses a serious burden of entry for \leedivdash{} experimenters.

\ViPErLEED{} drastically simplifies the user experience and expands the functionality and performance of existing solutions. The \python{} package controlling the computational aspects is named \calc{}. It manages the \leedivdash{} calculations --- acting as a wrapper for the established \tenserleed{} modules --- and contains higher-level logic and automation. By defining suitable defaults based on the theoretical structure and experimental data, \calc{} dramatically cuts down the required user input. Moreover, it provides additional functionality, such as optimization procedures not supported in the tensor-LEED formalism, automatic detection and preservation of surface and bulk symmetries during structural optimization, and acceleration of calculations through automated search procedures and parallelization.

\calc{} is designed with the vast scientific \python{} ecosystem in mind and leverages established numerical and scientific libraries such as \program{NumPy} \cite{numpy}, \program{SciPy} \cite{scipy}, \program{Matplotlib} \cite{matplotlib}, and \program{Scikit-learn} \cite{sklearn}. At the same time, it has an open application programming interface for future projects to exploit its novel features, such as the automated plane-group symmetry detection and structure symmetrization introduced in \Section{sec:symmetry}.

\section{Program Description}
As mentioned in \Section{sec:tensor_leed}, \ViPErLEED{} is based on the \tenserleed{} package \cite{tenserleed} used to perform the core \leedivdash{} calculations. These include the \quotes{full-dynamic} reference calculations, the calculation of amplitude changes \deltas{} as a function of parameter variations, as well as the search for an optimal combination of these variations. Input and output processing, as well as most high-level logic, are handled in a \python{} module named \calc{}, described in more detail below. Direct user interaction with the \tenserleed{} scripts is not required at any point. Some aspects of \tenserleed{} are also updated to a version 2.0 that is being released with, but independent of, the \calc{} \python{} package. The updated \tenserleed{} version is more computationally efficient and supports calculations on larger systems. For completeness, these changes are also briefly discussed in \Section{sec:tenser}. Finally, a graphical user interface (GUI) provides LEED pattern previews and can be used to generate input for the spot-tracking program \cite{viperleedSpot}.

The \calc{} package is optimized for UNIX-based systems (Linux, MacOS, and Windows subsystem for Linux -- WSL). Running natively on Windows is possible, but it involves some additional installation steps and configuration of the \fortran{} compilers. Installation instructions and details on setting up \fortran{} compilers can be found in the \ViPErLEED{} documentation available online \cite{viperleedOrg} or in the \SI{}.

\subsection{\ViPErLEED{} workflow\label{sec:calc_io}}
\leedivdash{} users wishing to work with existing solutions like \tenserleed{}, mostly based on the Van Hove--Tong codes \cite{vanhovetong1979}, are faced with several challenges. The first hurdle regards the input files: Specific input formats are expected, and these may be difficult for new users to understand. Specifically, standardized structure files, often obtained as a result of density-functional theory (DFT) calculations, cannot be used as input directly. Instead, they have to be translated to \quotes{LEED formats} manually, which is tiresome and error prone. The standard output files are also problematic. To the best of the authors' knowledge, none of the existing codes applying tensor-LEED calculations output the optimized structure in a standard crystal-structure format. As an example, \tenserleed{} optimization returns only a list of indices corresponding to the displacements of individual atoms that minimize the $R$ factor \cite{tenserleed}. Translating these raw indices back to atomic coordinates of the best-fit structure is, again, cumbersome and a likely source of mistakes. When multiple optimization steps are required, all of these steps must be performed by hand repeatedly. Lastly, users are often required to manually compile the source code at runtime. This is for example the case for \tenserleed{}: many array dimensions depend on the input data and \fortran{}~77 does not allow dynamic memory allocation.

\calc{} addresses these issues. It accepts input in a standard format and automatically translates it into the custom format required by \tenserleed{}. Likewise, \tenserleed{} outputs are translated back to the same standard formats by keeping track of the inputs and variations. \calc{} also removes the need for manual compilation at runtime. For each \tenserleed{} segment, it performs the steps shown in \Fig[b]{fig:workflow}: (i) Translate the user input into the format required by \tenserleed{}, (ii) compile the relevant \tenserleed{} code with appropriate array dimensions, (iii) execute the compiled code as a subprocess with adequate parallelization and with real-time monitoring if required, and (iv) process the \tenserleed{} output, producing user-readable output or input for the next \tenserleed{} segment. The compilation and execution of source code is a potential safety threat, as the source files may be modified with malicious intent. To ensure that only the expected \tenserleed{} code is executed, \calc{} contains hardcoded checksums for all expected \fortran{} files, and by default will only compile and run files conforming to these checksums.

Finally, \ViPErLEED{} also includes a  \quotes{bookkeeping} utility that keeps track of the performed calculations. It automatically updates a plain-text file (\param{history.info}) with information about the run segments, the resulting $R$ factors, and optional notes by the user. Additionally, it archives input and output files for every calculation into a \param{history} directory.

\subsection{Main \ViPErLEED{} input\label{sec:calc_input}}
The main input parameters defining \calc{} behavior are passed via the \param{PARAMETERS} input file. In the minimal case, this file defines which segments of \tenserleed{} should be executed and provides some information on how to interpret the structure files. However, the \param{PARAMETERS} file also allows in-depth control of \tenserleed{} calculation parameters and \ViPErLEED{} defaults. More information on these parameters and the choice of defaults will be given in \Section{sec:calc_defaults}. A comprehensive list of parameters and their default values is found in the \viperDoc{}.

Other than the general run parameters, the only input strictly required for calculating \iv{} curves is the structural data, that is, an atomic model for the diffracting surface. The chosen standard is the POSCAR format used by the Vienna \emph{ab-initio} simulation package (VASP) \cite{vasp}. This choice is based on the assumption that many \leedivdash{} calculations start from a DFT-optimized structure; POSCAR files can easily be generated by standard structure viewers such as VESTA \cite{vesta}. However, structural data can also be passed to \ViPErLEED{} directly from the atomic simulation environment (ASE) package \cite{ase}, which supports a variety of structural formats. Note that the surface is required to face the $+z$ direction, with the first two unit-cell vectors $\mathbf{a}$ and $\mathbf{b}$ in the surface $(x,y)$ plane. Additionally, \calc{} assumes the bottommost layers (lowest $z$ coordinates) to be bulklike. This means that symmetric slabs commonly used in DFT are not allowed. However, a \ViPErLEED{}-compatible slab can be obtained from a standard \quotes{DFT slab} by cutting below the \quotes{fixed} bulklike layers.

If structural optimization is to be performed, \calc{} requires two additional inputs: experimental \iv{} curves (file \param{EXPBEAMS.csv}), and instructions on optimization parameters and ranges (file \param{DISPLACEMENTS}). For \iv{} curves, \ViPErLEED{} uses standard CSV formatting for all input and output. The \param{DISPLACEMENTS} file defines the range over which each structural parameter is varied during tensor-LEED optimization, and can contain instructions about the order in which the parameters are optimized. Structural parameters that can be varied are (i) atom positions, (ii) vibration amplitudes, and (iii) site occupation (including changes of statistical chemical composition through, e.g., vacancies, dopants, or alloy constituents). For convenience, the \param{DISPLACEMENTS} file supports a variety of shorthand forms to address groups of atoms by element, layer, specific site type, or atom number. The exact syntax for the \param{DISPLACEMENTS} file is described in the \viperDoc{}. Plane-group symmetries of the surface are automatically preserved by \calc{} during displacements, such that symmetry-equivalent atoms in the input structure are kept symmetry equivalent during the optimization, unless the user explicitly reduces the symmetry by setting the desired plane group in the \param{PARAMETERS} file (see \Section{sec:symmetry}).

Besides the mandatory input discussed above, users can optionally provide additional files. Otherwise, these are automatically generated by \calc{}:
\begin{itemize}
\item First, an \param{IVBEAMS} file that specifies the \iv{} curves to be calculated. This can be automatically generated to match available experimental data. When no experimental data is given, \param{IVBEAMS} must be supplied by the user.
\item Second, a starting guess for vibration amplitudes \cite{foot_vib_amp} and site occupations. This information can be included in the \param{VIBROCC} file. Otherwise, sites are assumed to be \quotes{chemically pure}. If vibration amplitudes are not specified, an initial guess can be generated based on the Debye temperature $\Theta_\text{D}$ of the material and the experimental sample temperature $T$ --- both specified in \param{PARAMETERS} (see the \viperDoc{} for more details). In addition, a scaling factor may be defined for groups of atoms to reflect differences in the coordination environments. For example, undercoordinated atoms at the surface are expected to have a larger vibration amplitude. Since the vibration amplitudes themselves are parameters in the tensor-LEED optimization, this approximation is sufficient to obtain a reasonable starting point even when the Debye temperature is not known precisely.
\item Finally, energy-dependent phase shifts for electrons scattering on each type of atom in the structure. If not provided by the user (via a \param{PHASESHIFTS} file), these phase shifts will be generated automatically based on the input structure, using the \program{EEASiSSS} program \cite{rundgren2003}, distributed with the \ViPErLEED{} package with permission by the author.
\end{itemize}

\subsection{Main \ViPErLEED{} output}

\begin{figure*}[t]
\includegraphics[width=\pagefigure]{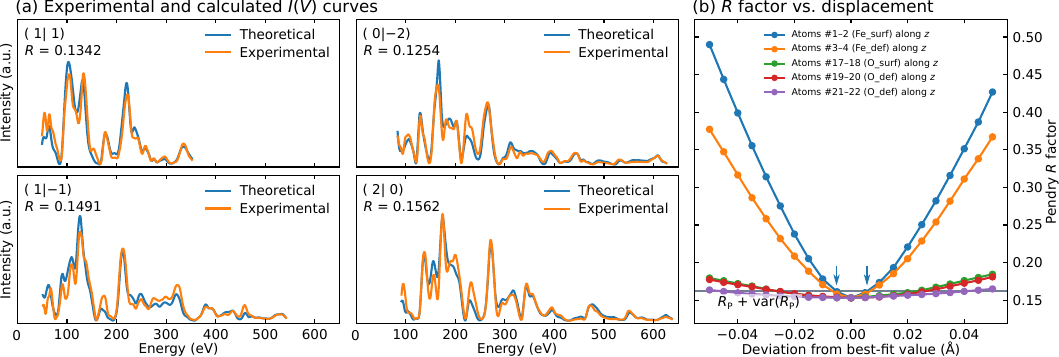}
\caption{\label{fig:fe2o3:select_beams}(a) Comparison of calculated to experimental \leedivdash{} curves for \hematiteOneby{}. All parts of the calculation have been performed using the \calc{} package. The figures are taken from the output files produced at the end of execution. Diffraction order and individual $R$ factors are indicated for each beam. (b) Error curves for the same system. They show the variation of the Pendry $R$ factor with respect to geometric out-of-plane displacements for the atoms in the topmost layer (cf.\ labels in \Fig{fig:fe2o3:layers}). The horizontal gray line indicates the minimum value of the Pendry $R$ factor plus its standard error. The intersection points of this line with an error curve (marked with vertical arrows for the blue curve) can be used as an estimate for the uncertainty of the corresponding parameter \cite{heinz2013book}.}
\end{figure*}

\ViPErLEED{} output files use the same formats as the input files. This ensures that, if further optimization steps are required, output from previous \ViPErLEED{} runs can be used directly as input for subsequent runs. The optimized structure is written to the files \param{POSCAR\_OUT} and \param{VIBROCC\_OUT}, and calculated \iv{} curves are again written in CSV format. In addition, \calc{} generates PDF files with plots comparing the calculated beams to experimental data, annotated with $R$ factors for each beam. Some plots as output by \calc{} for the \hematiteOneby{} example system are shown in \Fig{fig:fe2o3:select_beams}. More examples can be found in the \viperDoc{}. During structural optimization, a PDF file illustrating the search progress and state of convergence is generated in addition to the text-based log files (see Fig.~S1 of the \SI{}).

For diagnostic purposes, several supplementary files are written during each \calc{} run and stored in a directory called \param{SUPP}. These include the inputs and outputs of the \tenserleed{} programs, as well as modified \param{POSCAR} files generated during the structure analysis for troubleshooting.

\subsection{Structure analysis and dynamic defaults\label{sec:calc_defaults}}
As discussed in \Section{sec:calc_input}, one goal of \ViPErLEED{} is to reduce the number of parameters required as user input. While some \tenserleed{} parameters can simply be set to a default value that is normally appropriate, others must be chosen in accordance with the structure in question. However, in most cases, it is still possible to define case-specific defaults through educated guesses.

To give an example, the scattering phase shifts $\delta_\ell(E)$ depend on the angular momentum $\ell$. At large $\ell$, the phase shifts become small. Since the computing time strongly depends on the $\ell$ values (\Section{sec:tensor_leed}), a cutoff \lmax{} must be defined. \calc{} automatically selects the cutoff such that phase-shift values at $\ell > \ell_\mathrm{max}$ are smaller than a user-defined threshold (parameter \param{PHASESHIFT\_EPS}). Similar considerations apply to many parameters that can be determined from the structural input. For most low-to-medium-complexity cases, \calc{} requires user input of less than ten parameters for a standard optimization. A full list of parameters, their defaults, and descriptions on how they are determined is given in the \viperDoc{}, and some example inputs are included in the \SI{}. Here, the focus is on two non-trivial sets of parameters that \calc{} detects from the structural input: the repeat unit of the bulk, and plane-group symmetry.

\subsubsection{Detecting and constructing the bulk\label{sec:find_bulk}}
The shape of the \iv{} curves is strongly affected by (multiple) scattering events deep within the bulk, especially at high energies, where the mean free path of the electrons is long. The number of layers to be considered is determined dynamically during the full-dynamic LEED calculations. Hence, it is not sensible to provide slabs of fixed thickness as input; rather, the structural input should include a construction rule defining how to add additional bulk layers as required during the calculations. As discussed in \Section{sec:leed_theory}, this is implemented in \tenserleed{} through a bulk-doubling scheme \cite{tenserleed,moritzVanhove2022}. Two ingredients are needed: the repeat unit of the bulk, and a bulk repeat vector that, when applied iteratively, reconstructs the bulk structure. Since typical structure formats (like the \param{POSCAR} file) do not contain this information, it is necessary to either determine it automatically, or to include it explicitly as a user input. Both options are supported in \calc{}.


\paragraph*{Detecting bulk repeat.}
The user can manually define how many layers at the bottom of the slab constitute a bulk repeat unit (parameter \param{N\_BULK\_LAYERS}), as well as the bulk repeat vector (parameter \param{BULK\_REPEAT}, see \Fig{fig:fe2o3:layers}). However, \calc{} can also detect these automatically. It can determine the bulk repeat vector if the repeat unit of the bulk is explicitly provided (via \param{N\_BULK\_LAYERS}). It can even identify both if the slab contains at least two layers that are sufficiently bulklike. In this case, users only need to specify the limit below which the slab is bulklike (parameter \param{BULK\_LIKE\_BELOW}). To find a repeat unit, the program iteratively tests downwards translations that shift atoms of the $i$-th layer from the bottom to make them coincide with the position of the bottommost atom in the slab. Then, it compares the slabs before and after translation; if the atomic positions in the bulklike region are identical (within the tolerance given by parameter \param{SYMMETRY\_EPS}), the translation vector is accepted as a bulk repeat vector. Note that even when the repeat unit of the bulk is detected automatically as discussed above, the \param{N\_BULK\_LAYERS} and \param{BULK\_REPEAT} parameters are written to the \param{PARAMETERS} file explicitly. This causes \param{BULK\_LIKE\_BELOW} to be ignored in subsequent runs of \calc{}, ensuring that the repeat vector detected originally is preserved.

\paragraph*{Detecting surface superstructures.}
Another property that can be detected automatically in the initialization stage of the calculation is the relationship between the surface and bulk in-plane unit cells, that is, the periodicity of surface superstructures with respect to the bulk. Again, this relationship can be defined manually (parameter \param{SUPERLATTICE}, given either in Wood's notation or as a transformation matrix). It can also be detected automatically by checking all candidate in-plane translations, then choosing the smallest possible bulk unit cell. It is worth noting that when dealing with non-trivial surface-to-bulk relationships there may be multiple choices for the bulk unit cell and its orientation, which impact the numbering of diffraction spots. In such scenarios, users should ensure proper labeling of experimental beams. For this purpose, \calc{} generates an \param{experiment-symmetry.ini} file during initialization, containing details about detected unit cells and symmetry\todoHidden{Rename PatternInfo.tlm in code}. This file can be utilized as an input for the \ViPErLEED{} GUI to generate a \quotes{pattern file}. When employed with the \ViPErLEED{} spot tracker \cite{viperleedSpot}, this facilitates consistent beam labeling between experiment and calculations.

\paragraph*{Diagnostic files for bulk detection.}
\calc{} provides two diagnostic files to confirm the correct detection of the bulk: First, \param{POSCAR\_bulk}, containing the bulk unit cell constructed from \param{BULK\_REPEAT} and the two smallest unit vectors $\mathbf{a}_\text{bulk}$ and $\mathbf{b}_\text{bulk}$ in the surface plane; second, \param{POSCAR\_bulk\_appended}, which is the original \param{POSCAR} with additional bulk units added at the bottom using the \param{BULK\_REPEAT} vector. This allows a quick visual check of whether the detected bulk cell is aligned correctly with the slab and reproduces the bulk in all directions.

\subsubsection{Symmetry detection and structure symmetrization\label{sec:symmetry}}

\begin{figure*}[t]
\includegraphics[width=\pagefigure]{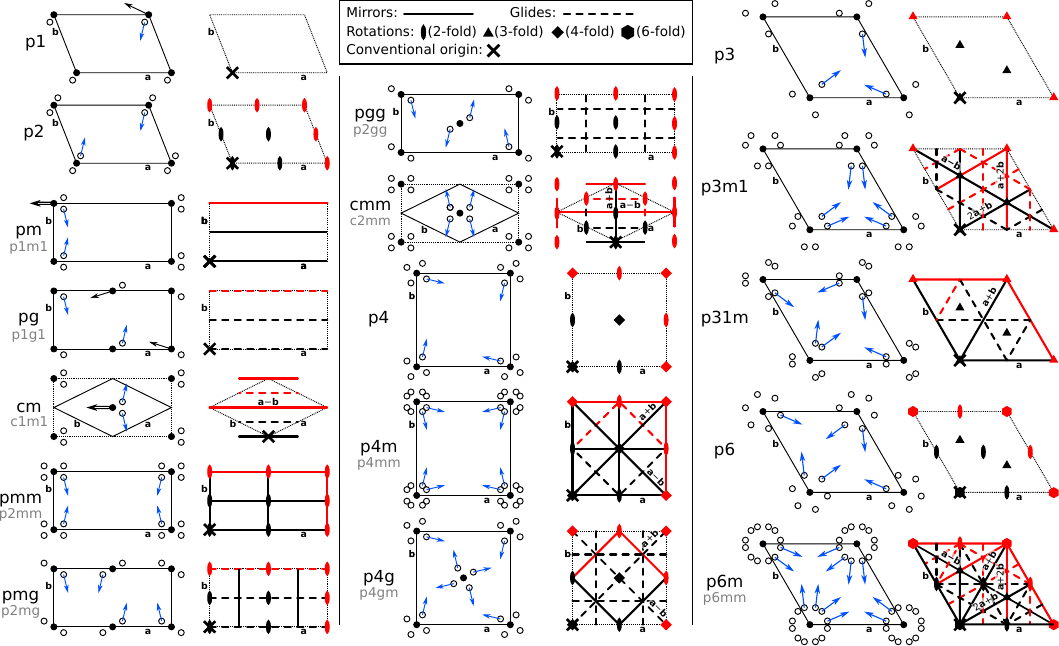}
\caption{\label{fig:plane_groups}The 17 plane symmetry groups. For each group, exemplary symmetry-equivalent atoms are shown as filled and open circles in the left panel, while the right panel shows the positions of the mirror, glide, and rotational symmetry elements. Symmetry elements drawn in red are related to those drawn in black via translations by integer multiples of the unit-cell vectors. Blue and black arrows indicate exemplary linked in-plane displacements. Double arrows in groups \emph{pm} and \emph{cm} indicate that movement is restricted along the mirror plane. Atoms on rotation axes cannot be displaced in plane. Full (four-character) plane group designations are indicated in gray, though \ViPErLEED{} uses short forms in all input and output.}
\end{figure*}

The symmetry of crystal surfaces can be one of the 17 plane symmetry groups (also called \quotes{wallpaper groups}) defined by the unit cell and rotation axes, as well as mirror and glide planes of the structure (see \Fig{fig:plane_groups}). This symmetry may already be known from DFT calculations, from scanning-probe measurements, or from a qualitative comparison of the LEED beams. It is normally desirable to preserve such a symmetry during structural optimization. Symmetry reduces the dimensionality of the parameter space, drastically speeding up the structure search. In current implementations of \leedivdash{} calculations, the user is required to identify symmetry-equivalent atoms, and to ensure they are displaced in a symmetry-equivalent manner during optimization. While this is relatively straightforward for simple symmetry groups, it carries a high potential for user error, for example, when threefold rotation axes are involved. \Figure{fig:plane_groups} shows how the wallpaper groups constrain in-plane displacements for symmetry-equivalent atoms. \calc{} provides fully automatic detection of the plane symmetry group and by default preserves the symmetry during optimization.

\paragraph*{Determining the minimal surface unit cell.}
To enforce some conventions and highlight the symmetry, \calc{} applies several automatic modifications to the unit cell in the structural input and output. First, unit cells are reduced to a minimum-circumference form while preserving the area, as shown in \Fig[a]{fig:ucell_movements}. This ensures that no higher-symmetry representation of a given structure is missed. An example is oblique cells that are reducible to a rectangular form: enforcing this transformation allows to find potential mirror planes. Second, for reasons of convention and simplicity, rhombic and hexagonal cells are expected in obtuse form (angle between $\mathbf{a}$ and $\mathbf{b}$ greater than 90$^\circ$). Acute unit cells are automatically transformed into obtuse while preserving handedness [\Fig[b]{fig:ucell_movements}]. Third, to avoid ambiguous beam labeling, the first unit-cell vector $\mathbf{a}_\text{bulk}$ of the bulk should always be taken as the shorter one. For example, for a rectangular bulk unit cell, a \woods{2}{1}{} supercell is closer to square than a \woods{1}{2}{} cell. However, this third point is not enforced, and \calc{} will only output a warning if the vectors are swapped.

\paragraph*{Detecting rotation axes, glide, and mirror planes.}
Once the unit-cell shape is set, \calc{} searches for rotation axes, glide planes, and mirror planes. The algorithm for finding the symmetry is briefly described in the following. First, it considers all rotational symmetries possible for the given unit-cell type (e.g., three- and six-fold rotations are only checked for hexagonal cells), and detects the highest-order rotation axes. Since only 2D operations are of interest, it is most efficient to use the sublayer \cite{foot_sublayer} with fewest atoms, 
as it yields the smallest set of candidate operations. Such candidates are then tested for the whole slab. If any rotation axes are found, the origin of the unit cell is shifted to the highest-order axis; if there are multiple highest-order rotation axes, the one closest to the original unit-cell origin is picked, as shown in \Fig[c]{fig:ucell_movements}. At this point, it is sufficient to check for mirror and glide planes going through the origin or through $(\frac{1}{4}, \frac{1}{4})$ of the unit cell. As seen in \Fig{fig:plane_groups}, this procedure uniquely determines the plane symmetry group.

\begin{figure}[t]
\includegraphics[width=\columnfigure]{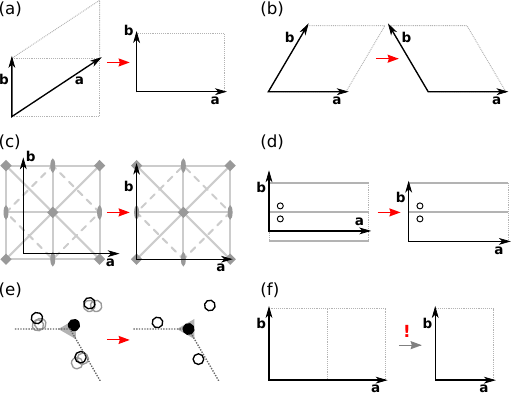}
\caption{\label{fig:ucell_movements}Modifications of the input unit cell during initialization. (a) Reduction to minimum-circumference form; (b) change from acute to obtuse form; (c) snapping of the origin to the closest highest-order rotation axis, shown on a \emph{p}4\emph{m} cell; (d) snapping of the origin to the closest mirror plane when there is no rotational symmetry, shown on a \emph{pm} cell; (e) symmetrization shown for a threefold rotation axis, where an atom close to the rotation axis (filled black circle) is snapped to the axis, and position errors between three nominally symmetry-equivalent atoms (open black circles, gray circles show rotated positions) are averaged such that they become truly equivalent; (f) reduction to a minimum-area surface unit cell. Steps (a)--(e) are applied automatically, while step (f) only produces a warning and outputs a suggested \param{POSCAR\_mincell} file.}
\end{figure}

If no rotation axis is found, mirror and glide symmetries are detected in analogy to rotational symmetries. In the minimum-circumference form of the unit cell, only the unit-cell-vector directions $\mathbf{a}$ and $\mathbf{b}$ and the diagonals $\mathbf{a}+\mathbf{b}$ and $\mathbf{a}-\mathbf{b}$ (plus all 30$^\circ$-rotated directions $2\mathbf{a}+\mathbf{b}$, etc.\ for hexagonal cells) need to be considered to find all possible mirror/glide planes \cite{foot_symmetry}. The unit-cell origin is again shifted (by the smallest possible distance) to coincide with a mirror plane, or with a glide plane if no mirror plane was found, as shown in \Fig[d]{fig:ucell_movements}. In cases where the orientation of the mirror/glide planes is ambiguous (groups \emph{pm}, \emph{pg}, \emph{cm} and \emph{pmg}), the symmetry can be fully specified by indicating the direction of the mirror or glide plane through the origin. For example, \emph{pm}$[1~0]$ indicates \emph{pm} symmetry with mirror planes parallel to the $\mathbf{a}$ axis. The identified plane group symmetry is printed to the \calc{} log file and added as a comment in the \param{POSCAR} file.

\paragraph*{Symmetrization.}
Once the symmetry has been determined, \calc{} performs a symmetrization step to correct minor errors potentially introduced, for example, by not-fully-converged DFT relaxations. Positions of atoms that are equivalent within a given tolerance (parameter \param{SYMMETRY\_EPS}) are averaged such that they are equivalent within floating-point precision. Atoms that should be locked to a rotation axis or mirror plane snap to that position, as shown in \Fig[e]{fig:ucell_movements}. This can be helpful since \leedivdash{} calculations are sensitive to position deviations in the picometre range.

The applied symmetry group can also be reduced manually, for example when the surface is suspected to possess a lower symmetry than the initial input structure. The \param{SYMMETRY\_FIX} parameter allows to manually set any plane symmetry subgroup of the automatically detected group, including \emph{p}1 if symmetry should be turned off entirely. In addition to the \param{SYMMETRY\_FIX} parameter, which globally modifies the symmetry to be used, there is also a more local option to deactivate or reduce symmetry only for specific operations in the \param{DISPLACEMENTS} file with the \param{SYM\_DELTA} tag. However, since this will result in breaking the symmetry in subsequent runs, users are discouraged from utilizing the \param{SYM\_DELTA} tag except for highly specific cases. More details on both types of symmetry reductions are found in the \viperDoc{}.

\paragraph*{Detecting beam equivalence and beam assignment.}
Superstructures often exhibit different symmetry than the underlying bulk. Hence, single terraces of a sample may feature multiple \quotes{patches} of the same superstructure mirrored or rotated relative to one another, as induced by the plane group of the bulk. Additionally, different terrace orientations may be present when the bulk possesses screw or glide symmetry (some well-known cases are, e.g., Si(001) \cite{si_review} or Fe$_3$O$_4$(001) \cite{bliem2014}). This induces corresponding rotation or mirroring of the superstructures.

Since symmetry-induced domains normally have sizes larger than the typical coherence length of the electron beam, they contribute to the diffraction pattern incoherently. Such effect must be considered to correctly match calculated and experimental \iv{} curves. \calc{} dramatically simplifies the user experience in this respect: instead of requiring the user to specify which calculated beams contribute to each experimental beam and should be averaged together, it detects beam equivalence automatically, assuming a statistical distribution of symmetry-induced domains. It also does not require the user to manually assign calculated to experimental beams: it does so automatically based on the beam labels [e.g., $(1~0)$ is the first-order diffraction spot in the direction of the first reciprocal unit-cell vector of the bulk, $\mathbf{a}^*_\text{bulk}$]. The automatic beam averaging is enabled by \calc{}'s capability to detect symmetry (\Section{sec:symmetry}). The same algorithm used for the whole slab is applied to the bulk, extending the search also to glide planes and screw axes with translation vectors parallel to the surface normal.

\subsection{\label{sec:parallelization}Automation and parallelization}
A typical \leedivdash{} optimization with \tenserleed{} proceeds as follows: (i) execution of a full-dynamic calculation for the starting guess structure; (ii) optimization along a given displacement direction; (iii) further optimizations along other directions. [Steps (ii) and (iii) are distinct because the \tenserleed{} search module only supports 1D displacements for atom positions, such that multiple executions are required to fully optimize 3D atomic positions.] Steps (i) and (iii) are repeated until convergence. In \tenserleed{}, the user must perform all of these steps manually. This becomes especially cumbersome as one segment's output must be manually translated into an input for the next segment. \calc{} simplifies the workflow. First, it handles the input/output processing internally. Second, it allows arbitrary queuing of segments with the \param{RUN} parameter. Finally, it allows to specify multiple sets of displacements in the \param{DISPLACEMENTS} file, such that they are automatically executed in the requested order. These optimizations can be looped until the $R$ factor converges, allowing for an autonomous search in a wide parameter space. The \viperDoc{} gives more details on the \param{RUN} parameter and \param{DISPLACEMENTS} file. As described in Section~S1 of the \SI{}, \calc{} also speeds up each optimization run in \tenserleed{} by monitoring the search progress and adjusting convergence parameters accordingly.

Besides automating the optimization workflow, \calc{} also enhances the parallelization of \leedivdash{} calculations. The parallelization offered by the native \tenserleed{} code is limited to the \quotes{search} module. \calc{} introduces additional parallelization at the \python{} level. Parallelization is applied to the full-dynamic calculation by assigning individual energies to separate processes. This is possible since each energy step can be calculated independently from the others. This approach enables further acceleration through a dynamic use of \lmax{} at each energy: At low energies, where phase shifts of large angular momenta are negligible, small \lmax{} values are used. Parallelization is also applied to the \deltas{} calculation. Here, the atoms under variation are independent and can be calculated in a parallel manner. Note that, since the single calculations are vastly more computationally expensive than the overhead, the parallelization offered by \calc{} comes close to the efficiency theoretically achievable with parallelization directly in the \fortran{} code.

\subsection{Advanced optimization}

\subsubsection{Mixed-termination surfaces\label{sec:domains}}
Many systems can sustain coexisting surface reconstructions. In such cases, the intensities measured experimentally originate from an incoherent superposition of the diffracted beams from each termination. Some packages (like \tenserleed{} \cite{tenserleed} and MSATLEED \cite{vanhove_web}) already implement optimization for this scenario.

\calc{} provides a simplified interface to the multi-domain optimization implemented in \tenserleed{}. A multi-domain calculation can be started either by supplying paths to the input structures or by providing pre-existing reference calculations (and corresponding \param{Tensor} files). In the latter case, \calc{} automatically performs a consistency check for compatibility of the input files. If the consistency check fails, \calc{} determines which parameters are shared between the different structures, harmonizes the input, and performs new, mutually compatible, full-dynamic calculations. Structural optimization is then performed using the tensor-LEED approximation including the fractional coverage of the domains as an additional parameter.

\subsubsection{Full-dynamic optimization}
\calc{} also provides an option for optimizing parameters that are inaccessible to the tensor-LEED approximation. Examples are the incidence angles $\theta$ (polar) and $\phi$ (azimuthal) of the primary electron beam, the imaginary part of the inner potential $V_{0\text{i}}$, and the lattice parameters of the input unit cell. For the latter, a scaling factor can be applied to one of the unit-cell vectors, to both in-plane unit-cell vectors at once, or to all three at once (i.e., scaling the volume of the unit cell).

Optimization of these parameters is enabled through repeated full-dynamic calculations. A \quotes{full-dynamic optimization} is set up using the \param{OPTIMIZE} parameter. It supports defining the parameter to optimize, the initial step size of variation, and convergence criteria. It is important to note that this kind of full-dynamic optimization is computationally significantly more expensive than tensor-LEED-based optimizations.



\subsection{\label{sec:tenser}TensErLEED version 2.0}  
Although \ViPErLEED{} is backwards compatible with the most recent version of \tenserleed{} at the time of writing (version 1.6), it was expedient during development to also make some improvements on the \tenserleed{} codebase. This resulted in a new version 2.0 of \tenserleed{} \cite{tenserleedRepo}, which is released with \ViPErLEED{} but can also be used independently. Changes include some minor bug fixes, increased precision of some computational variables, and improved computational performance, especially of the full-dynamic \quotes{reference} calculation. The speed-ups mainly come from replacing custom matrix-algebra routines by better-optimized BLAS and LAPACK versions \cite{lapack}, skipping redundant calculations, and refactoring certain routines for more efficient memory access.

The user interface (i.e., format of the input and output files) in \tenserleed{} 2.0 was modified to extend the size limits of systems that can be calculated. Previous \tenserleed{} versions severely limited the size of the unit cell, as well as related quantities such as the number of diffracted beams to calculate. The \tenserleed{} 2.0 input now allows larger values. Furthermore, some variables that were obsolete in the \tenserleed{} code but were still required in the input were removed, and others (e.g., imaginary part of the inner potential \param{VPI} and the filament work function \param{WORKFN}) were moved from hardcoded values in a runtime-created \fortran{} script (\param{muftin.f}) to standard input parameters.

Modifications to the \tenserleed{} code by the Solid-State Physics group in Erlangen and the \ViPErLEED{} team are released together with \ViPErLEED{} under the terms of the GNU General Public License version 3 (or any later version) \cite{gplv3}.

\subsection{Ongoing development}
The \ViPErLEED{} version 1.0 released with this work is a fully functional \todoHidden{[@michele: fish back working GUI code for master]} open-source package, licensed under the GNU General Public License version~3 or later \cite{gplv3}. Further development is in progress and will be published on the GitHub repository \cite{viperleedRepo}.

Already in version 1.0, \ViPErLEED{} can start calculations based on structure input in the form of ASE \param{Atoms} objects \cite{ase}. This interface will be expanded to allow full control of all \ViPErLEED{} features. This should, for the first time, allow easy high-throughput \leedivdash{} calculations for automatically generated surface structures. Many computational groups in the field are currently striving to solve complex surface structures without relying on human intuition \cite{noordhoek2024,li2023,musa2022,karsten2023,wanzenboeck2022,merte2022,behler2017}. As demonstrated in \RefInline{primorac2016}, the $R$ factor can be used as a cost function for guiding machine-learned structure prediction. The sensitive feedback provided by \leediv{} may provide an attractive and new use of the technique for structure search in conjunction with total energies yielded by DFT calculations. A major benefit of using \leediv{} for structure determination is its direct link to experimental data.

Another major change currently under development concerns a replacement for the \tenserleed{} optimization (\quotes{search}) algorithm. While \ViPErLEED{} will continue to use the \tenserleed{} code for the full-dynamic calculations and calculations of amplitude changes \deltas{}, the \tenserleed{} algorithm for finding the combination of \deltas{} that minimizes the $R$ factor is somewhat dated. In the future, the current one-dimensional, grid-bound search may be replaced by a continuous search by calculating \deltas{} values on demand. This would allow direct application of, and free choice among established global optimization algorithms available from many optimized \python{} packages (e.g., \program{SciPy} \cite{scipy}, \program{Scikit-Optimze} \cite{skopt}, \program{Keras} \cite{keras}, \program{Ray} \cite{ray},  \program{JAX} \cite{jax}). Most of these modern evolutionary or (stochastic) gradient-based algorithms are likely to be much more effective at optimizing the structure than the original custom algorithm implemented in \tenserleed{}.

\section{\boldmath{}Benchmark result: The \woods{1}{1}{} termination of \hematite{}}
The \ViPErLEED{} documentation (see \SI{}) guides the user through the application of \ViPErLEED{} on a few example systems [among others, Ag(100), Ni(110), and Ir(100)-\woods{2}{1}{}O]. This section outlines the main results obtained with \ViPErLEED{} on a \quotes{medium-complexity} oxide surface: The \woods{1}{1}{} termination of $(1\overline{1}02)$-oriented hematite ($\alpha$-Fe$_2$O$_3$) \cite{henderson2002,lad1988,tanwar2007}. An introduction to the system and details on sample preparation are given in the \SI{}.

To ensure adequate sample conductivity of this wide-bandgap semiconducting material, \leedivdash{} data were acquired on a slightly Ti-doped (0.03~at.\%) \hematite{} film (see Section~S2 in the \SI{} for details on the growth). The Ti-doped films expose atomically flat, single-crystalline surfaces; their \woods{1}{1}{} and \woods{2}{1}{} structures are identical to those of undoped crystals \cite{franceschi2020,kraushofer2018}. The \woods{1}{1}{} termination was prepared through repeated cycles of sputtering (10~min at 1~keV, $5\times10^{-7}$~mbar Ar, $\approx65$~nA/mm$^2$) and annealing (30~min at 550\degC{}, $7\times10^{-7}$~mbar O$_2$, 20\degC{}/min ramp rate). Before each LEED measurement, the sample was heated for 10--20~min at $\approx300$\degC{} to desorb H$_2$O \cite{jakub2019}. The cleanliness and quality of the surface was ensured by x-ray photoelectron spectroscopy and scanning tunneling microscopy.

LEED measurements were acquired at room temperature with an Omicron SpectaLEED (LaB$_6$ cathode, three grids, one microchannel plate), and a DMK 33GX265 camera (The Imaging Source Europe GmbH, 80~ms exposure, 0~dB gain, 15-frames averaging, $3\times3$ binning), using an electron energy range of 30--750~eV ($0.5$~eV steps). The extracted \iv{} data span a total energy range of $10.8$~keV after averaging symmetry-equivalent beams. Flat-field correction was performed by acquiring reference videos on the sample plate (see \viperSpot{} for more details). Intensities were normalized to the energy-dependent net beam current, $I_0$, corrected for the energy-dependent current drawn by the LEED electronics, $I_{00}$. To minimize artifacts from the LEED grids, the data were averaged from two sets of measurements acquired at sample--grid distances different by 2~mm (more details in \viperSpot{}). The data were taken during night to avoid magnetic-field fluctuations induced by streetcars traveling in front of the building. Static magnetic fields were compensated with the help of two coils mounted outside the vacuum chamber (see also \viperMeas{}). Since the sample has $pg$ plane-group symmetry, the LEED intensities at perpendicular incidence have mirror symmetry in only one direction and no rotational symmetry. The crystal was rotated around the $[\overline{1}101]$ in-plane direction to ensure that the incident beam lies in the mirror plane of the LEED pattern, i.e., to achieve identical \iv{} curves of the $(h|k)$ and $(\overline{h}|k)$ spots. This ensures that the incidence plane is perpendicular to the surface. In turn, this allows to fix the azimuthal angle of incidence in the calculations, such that only the polar one requires fitting.

\begin{figure}
\includegraphics[width=\columnfigure]{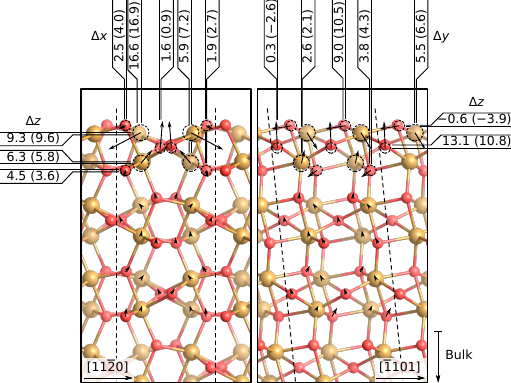}
\caption{\label{fig:fe2o3:leed_relaxed}Structure of the \hematiteOneby{} surface resulting from \leedivdash{} optimization. Oxygen atoms are drawn in red, iron atoms in yellow. The left and right panels show side views along the $[\overline{1}101]$ and $[\overline{1}\overline{1}20]$ in-plane directions, respectively. Dashed lines delimit one surface unit cell. Arrows indicate the displacements of atoms relative to the bulk-truncated \hematite{} surface, and are exaggerated by a factor of ten. The bulk-truncated positions of atoms in the surface layer are represented as dashed circles. The $x$, $y$, and $z$ components of the displacements of the atoms in the topmost surface layer are indicated (in picometres). Numbers in parentheses refer to the corresponding displacements as obtained by DFT. Table~S1 of the \SI{} reports the displacement of all atoms, as well as their uncertainties, derived from \leediv{}. There, all the DFT displacements are also listed. See Section~S3 of the \SI{} for more details on the DFT calculations.}
\end{figure}

\Figure{fig:fe2o3:leed_relaxed} summarizes the outcome of a \leedivdash{} analysis performed with \ViPErLEED{} on \hematiteOneby{}. The \SI{} includes the raw LEED data as well as final output files. It also describes in detail the steps leading to the optimization, as well as the parameters varied. \Figure{fig:fe2o3:leed_relaxed} shows two side views of the system, including values of atomic displacements predicted by the relaxed, lowest-energy DFT model (in parentheses), and by \leedivdash{} optimization. \Figure[a]{fig:fe2o3:select_beams} shows exemplary best-fit \leedivdash{} spectra to demonstrate the fit quality. All the experimental and calculated \iv{} curves are plotted in Fig.~S3 of the \SI{}. The best-fit model of the structural analysis reproduces the measured \leedivdash{} spectra with an overall Pendry $R$ factor of $R_\text{P} = 0.154$. For educational purposes, \leedivdash{} optimization was started from a bulk-truncated slab \cite{maslen1994} rather than the DFT-optimized model. The relaxations obtained by \ViPErLEED{} and DFT are very close (within $3.5$~pm), and consistent within the uncertainty estimates that can be derived from \leediv{} (see also Table~S1 of the \SI{}). The largest relaxations involve atoms of the topmost layer (cf.\ displacements marked in \Fig{fig:fe2o3:leed_relaxed}). At the surface, the Fe atoms relax downwards into the plane of the four surrounding O atoms, while O atoms from the first subsurface \quotes{sublayer} move closer to the surface. Effectively, this moves the Fe atoms of the topmost cation layer to coincide with the plane of their equatorial O atoms (distances from the plane decreased from $\approx21$~pm to $\approx3$~pm).


\begin{table}[t]
\caption{\label{tab:exec_time}Timing benchmarks for the full-dynamic \quotes{reference} calculation performed for \hematiteOneby{}.}
\begin{tabular*}{\columnfigure}{@{\extracolsep{\fill}}@{}lrcc@{}}%
\toprule
Setup\footnote{Calculations were performed on the Vienna scientific cluster (VSC-4 \cite{vsc}) on a node with two Intel\textregistered{} Xeon\textregistered{} Platinum 8174 processors, with a total of 96 logical cores, and 96~GB memory. The \param{ifort} compiler (version 2021.7.1) from the Intel\textregistered{} oneAPI\textregistered{} high-performance computing packages was used. All calculations used a maximum of two threads for LAPACK parallelization, by setting the oneAPI\textregistered{} math-kernel library \param{MKL\_NUM\_THREADS} environment variable.}%
  &%
  \lmax{}\footnote{Maximum angular momentum quantum number used for the expansion in spherical harmonics, as described in \Section{sec:tensor_leed}.}
  &%
  \param{N\_CORES}\footnote{Number of logical cores used for running calculations at different energies in parallel. See also \Section{sec:parallelization}.}
  &%
  Execution time (min)\\
\midrule
\tenserleed{} v1.6.0 & 12\hphantom{\textsubscript{e}} & 1  & 89\\
\tenserleed{} v2.0.0 & 12\hphantom{\textsubscript{e}} & 1  & 69\\
\ViPErLEED{}\footnote{\label{tab_fn:tenser}Using \tenserleed{} v2.0.0.} & 6--12\footnote{\label{tab_fn:dynamic_lmax}Dynamically assigned by \ViPErLEED{} with the default settings for the \param{LMAX} parameter. The same cutoff for phase shifts was employed in the calculation with dynamic \lmax{} as in the ones with fixed \lmax{}, but was evaluated independently at each energy, rather than using the \lmax{} determined at the highest beam energy for all calculations.} & 1  & 63\\
\ViPErLEED{}\textsuperscript{\ref{tab_fn:tenser}} & 6--12\textsuperscript{\ref{tab_fn:dynamic_lmax}} & 48 & $2.5$\\
\bottomrule
\end{tabular*}
\end{table}

\Table{tab:exec_time} summarizes the duration of the full-dynamic calculations for the \woods{1}{1}{} termination of \hematite{} reported in \Fig{fig:fe2o3:leed_relaxed}, as obtained with different setups. Through efficient use of parallelization, \ViPErLEED{} produces results in less than 3~min, more than 30 times faster than the original \tenserleed{} configuration. This significant reduction of computation time not only enhances efficiency but also unlocks new opportunities for exploring larger, more complex, and previously inaccessible systems. More importantly, while harder to quantify, the time needed to set up this calculation with \ViPErLEED{} is also on the order of minutes. In contrast, translating the structural and experimental data to \tenserleed{} input would be much more time intensive and error prone. For inexperienced users in particular, setting up the \tenserleed{} input for the first time can be daunting, given the numerous parameters with non-telling names that need definition. When the input contains mistakes, the \fortran{} modules crash, often without explanation. Instead, \ViPErLEED{} requires only a handful of well-documented parameters. Moreover, it provides sensible feedback when executed, consisting of both fine-grained error messages, warnings based on a large number of \quotes{sanity checks}, as well as informative diagnostic files.

\section{Conclusions}
This work describes a new user-friendly and flexible open-source package for \leedivdash{} calculations and structural optimization, built on top and as an extension of the well-established \tenserleed{} package. This is part of the larger \ViPErLEED{} project that also addresses data acquisition \cite{viperleedMeasurement} and extraction \cite{viperleedSpot}. Both the barrier of entry for new users and the potential for human error are significantly reduced by an expedient use of default parameters. The defaults are either static or automatically derived from properties of the experimental data and surface structure. Most significantly, surface symmetries are automatically detected and preserved during optimization. The program also automatically detects how calculated beams should be averaged and matched to an experimental beam set. All segments of \leedivdash{} calculations run under one combined user interface with standard input and output formats. Additional developments include simple optimization procedures for parameters inaccessible to the tensor-LEED approach, such as the incidence angle of the electron beam or the imaginary part of the inner potential. Computational performance of the \tenserleed{} package was improved both via top-level parallelization of the full-dynamic \quotes{reference} calculations and calculations of \quotes{delta amplitudes} \deltas{}, and through optimization directly in the \tenserleed{} code. This gain in speed and accuracy (in the sense of freedom from human errors) can be leveraged to perform LEED analysis of large systems that were previously impractical to tackle, as exemplified by a recent work on a surface-telluride phase with \woods{10}{10}{} periodicity on Pt(111) \cite{kisslinger2023}.

\begin{acknowledgments}
This research was funded in part by the Austrian Science Fund (FWF) under \href{https://doi.org/10.55776/F81}{doi:10.55776/F81}, Taming Complexity in
Materials Modeling (TACO), and  by the European Research Council (ERC) under the European Union’s Horizon 2020 research and innovation programme (grant agreement No. 883395, Advanced Research Grant ‘WatFun’). For the purpose of open access, the authors have applied a CC BY public copyright licence to any Author Accepted Manuscript version arising from this submission.

Part of the \leedivdash{} calculations were performed on the Vienna Scientific Cluster (VSC).

The authors extend their acknowledgments to all beta testers for reporting issues, and in particular to Jascha Bahlmann, Maximilian Buchta, Holger Diem, Julian Hochhaus, Jan Lachnitt, Matthias Meier, Simon Moser, Josef Myslive\v{c}ek, Samuel Petrov, Karl-Michael Schindler, Alexander Schneider, and Martin Setv\'{i}n for their valuable feedback. Furthermore, Matthias Blatnik, Jes\'{u}s Carrete Monta\~{n}a, Ralf Wanzenb\"{o}ck, Florian Buchner, Georg Madsen, Florian Libisch, and Christoph Schattauer are thanked for fruitful discussions.

Finally, the authors are grateful for the permission to redistribute as part of \ViPErLEED{} the \program{EEASiSSS} program for calculating phase shifts (John Rundgren) \cite{rundgren2003,rundgren2007,rundgren2021} and its dependencies (Eric Shirley) \cite{shirley1991}, as well as \tenserleed{} (Volker Blum and Klaus Heinz) \cite{tenserleed}. Volker Blum, Klaus Heinz, and Eric Shirley are acknowledged for relicensing their contributions to \tenserleed{} and \program{EEASiSSS} under the terms of the GNU GPLv3 (or later).


\end{acknowledgments}

\providecommand{\enquote}[1]{#1}

\begin{thebibliography}{103}
\bibitem [{foo({\natexlab{a}})}]{foot_info_in_iv}%
  \BibitemOpen
  \bibinfo {note} {The multiple-scattering nature of electron--solid
  interaction is the main source of the high information density in
  \leedivdash{} data compared to, for example, surface-truncation rods in x-ray
  scattering.}\BibitemShut {Stop}%
\bibitem [{foo({\natexlab{c}})}]{foot_vib_amp}%
  \BibitemOpen
  \bibinfo {note} {The vibration amplitude as defined here (i.e., the square
  root of the $\langle{}u^2\rangle_T$ value at temperature $T$) is based on the
  isotropic average of the squared 3D vibration displacements. These values may
  also include static disorder, such as small, random displacements due to
  defects or adsorbates.}\BibitemShut {Stop}%
\bibitem [{foo({\natexlab{d}})}]{foot_sublayer}%
  \BibitemOpen
  \bibinfo {note} {A \quotes{sublayer} is defined as a set of chemically
  equivalent atoms with the same $z$ coordinate (within tolerance) to
  distinguish it from the notion of \quotes{layer} as used in \tenserleed{}:
  the latter is a composite object that may often be several \aa{}ngstr\"{o}ms
  thick. A \quotes{layer} is commonly composed of several
  \quotes{sublayers}.}\BibitemShut {Stop}%
\bibitem [{foo({\natexlab{e}})}]{foot_symmetry}%
  \BibitemOpen
  \bibinfo {note} {Note that this approach implies that the correct symmetry
  may not be found if the \param{POSCAR} unit cell is not primitive in the
  surface plane. However, {\calc{}} will automatically recognize non-primitive
  unit cells, output a warning, and write a \param{POSCAR\_mincell} file
  containing a slab with a primitive unit cell.}\BibitemShut {Stop}%
\end{thebibliography}
%

\end{document}